\documentclass[aps,prd,preprint,superscriptaddress]{revtex4}

\usepackage{graphicx}
\usepackage{amsmath,amssymb}
\usepackage{bm}
\usepackage{color}

\def\ee{\end{eqnarray}}

\def\p{\partial}

\def\=:{=\hspace{-.7em}\raisebox{1.1ex}{.}\hspace{.1em}\raisebox{-0.2ex}{.} }

\def\ee{\end{eqnarray}}

\def\p{\partial}

\def\=:{=\hspace{-.7em}\raisebox{1.1ex}{.}\hspace{.1em}\raisebox{-0.2ex}{.} }

\newcommand {\beq}{\begin{eqnarray}}
\newcommand {\eeq}{\end{eqnarray}}
\newcommand {\non}{\nonumber\\}

\newcommand {\1}[1]{\frac{1}{#1}}

\newcommand {\ph}{\varphi}

\newcommand {\del}{\partial}

\newcommand {\tr}{{\rm tr}\,}

\begin{document}

\vspace*{-2cm}
\begin{flushright}
{\tt YGHP-15-01}
\end{flushright}

\title{Non-Abelian Sine-Gordon Solitons:\\
Correspondence between 
$SU(N)$ Skyrmions 
and ${\mathbb C}P^{N-1}$ Lumps
}


\author{Minoru Eto}
\affiliation{Department of Physics, Yamagata University, Kojirakawa-machi 1-4-12, Yamagata, Yamagata 990-8560, Japan}

\author{Muneto Nitta}

\affiliation{
Department of Physics, and Research and Education Center for Natural 
Sciences, Keio University, Hiyoshi 4-1-1, Yokohama, Kanagawa 223-8521, Japan\\
}


\date{\today}
\begin{abstract}
Topologically stable 
non-Abelian sine-Gordon solitons have been found recently 
in the $U(N)$ chiral Lagrangian 
and 
a $U(N)$ gauge theory 
with two $N$ by $N$ complex scalar fields 
coupled to each other. 
We construct the effective theory on 
a non-Abelian sine-Gordon soliton  
that is a nonlinear sigma model 
with the target space ${\mathbb R} \times {\mathbb C}P^{N-1}$.
We then show that 
${\mathbb C}P^{N-1}$ lumps on it 
represent $SU(N)$ Skyrmions in the bulk point of view, 
providing a physical realization of 
the rational map Ansatz for Skyrmions 
of the translational (Donaldson) type.
We find therefore that Skyrmions 
can exist stably without the Skyrme term.

\end{abstract}

\maketitle

\section{Introduction}

When a soliton equation is integrable, 
one can construct exact analytic solutions in principle.
Among topological solitons and instantons,
Yang-Mills instantons \cite{Belavin:1975fg} and 
Bogomol'nyi-Prasad-Sommerfield (BPS) 
monopoles 
\cite{Bogomolny:1975de,Prasad:1975kr} 
are such examples studied in detail 
both in 
physics and mathematics, 
for which exact solutions are accessible from 
the Atiyah-Drinfeld-Hitchin-Manin 
\cite{Atiyah:1978ri} 
and Nahm \cite{Nahm:1979yw}
constructions, respectively. 
For BPS monopoles, 
Donaldson proposed a rational map construction 
\cite{Donaldson:1985id},
in which three-dimensional space 
is decomposed into one particular direction 
and its orthogonal plane is
parametrized by a complex coordinate. 
Recently, a physical interpretation of the 
Donaldson's rational map was 
provided in Ref.~\cite{Nitta:2010nd} 
by putting monopoles into the Higgs phase,
in which vortices that confine monopoles 
extend to the above-mentioned one particular 
direction.
A spherical rational map construction 
was also proposed in Ref.~\cite{Jarvis:1998} in which 
three-dimensional space 
is decomposed into a sphere and a radial direction.

The Skyrme model that describes baryons as solitons 
known as Skyrmions 
\cite{Skyrme:1962vh} 
is not integrable, 
unlike its BPS version proposed recently 
whose Lagrangian 
consists of only a six-derivative term
and a potential term \cite{Adam:2010fg}. 
Since exact solutions are impossible to obtain 
for the original Skyrme model,
approximate analytic solutions are 
the most useful if they exist, 
unless one obtains solutions numerically.
One such approximation is
the Atiyah-Manton Ansatz 
\cite{Atiyah:1989dq,Atiyah:1992if} 
in which an approximate Skyrme field is obtained from 
a holonomy of a Yang-Mills instanton configuration 
integrated along one particular direction.
A physical realization of 
the Atiyah-Manton ansatz 
has been obtained recently
\cite{Eto:2005cc} 
in which a Skyrmion is realized as 
a Yang-Mills instanton absorbed into 
a domain wall that is placed perpendicular to 
the above-mentioned one particular direction.
The other more useful approximation is 
the rational map Ansatz proposed in Refs.~\cite{Houghton:1997kg,Manton:2004tk},
in which three-dimensional space 
is decomposed into a sphere and a radial direction, 
as for the Jarvis's spherical 
rational map Ansatz for BPS monopoles. 
This Ansatz was also generalized to 
$SU(N)$ Skyrmions
\cite{Ioannidou:1999mf}.
For a recent application to 
realistic situation, see Ref.~\cite{Lau:2014baa}. 
While this Ansatz gives only an initial configuration 
for numerical relaxation, a physical realization 
of this Ansatz can be also given 
as a spherical domain wall 
\cite{Gudnason:2013qba,Gudnason:2014gla} 
which can be stabilized in a Skyrme model 
with a six-derivative term. 
On the other hand, a Donaldson-type rational map 
Ansatz for Skyrmions has been found  
\cite{Nitta:2012wi,Nitta:2012rq,Gudnason:2014nba,
Gudnason:2014hsa} 
together with its physical realization  
in which ${\mathbb C}P^1$ lumps inside a domain wall 
are Skyrmions in the bulk 
in the Skyrme model with the modified mass term 
admitting two discrete vacua 
\cite{Kudryavtsev:1999zm}.
However, a generalization to $SU(N)$ Skyrmions
has a difficulty
that such a potential term admitting a 
domain wall is not known.

The purpose of this paper is 
to give a physical realization of 
an $SU(N)$ rational map of the Donaldson type 
for Skyrmions. 
A key ingredient is 
a non-Abelian sine-Gordon soliton 
proposed recently \cite{Nitta:2014rxa} (see also earlier work \cite{Gepner:1984au}),
in which it has been found that 
a $U(N)$ chiral Lagrangian 
with the usual pion mass term, instead of $SU(N)$, 
admits a topologically stable 
non-Abelian sine-Gordon soliton.
The point is that 
the $U(N)$ group has the structure of
$[SU(N) \times U(1)] /{\mathbb Z}_N$, and 
consequently there exists 
a topologically nontrivial closed path winding 
around the $U(1)$ group $1/N$ times 
together with an $SU(N)$ path from the unit element 
to an element in the center ${\mathbb Z}_N$.
The diagonal $SU(N)$ symmetry in the vacuum  
is spontaneously broken 
into an $SU(N-1) \times U(1)$ subgroup 
in the presence of the non-Abelian 
sine-Gordon soliton, 
giving rise to localized  
${\mathbb C}P^{N-1} \simeq 
SU(N)/[SU(N-1) \times U(1)]$ 
Nambu-Goldstone modes.
Therefore, the term ``non-Abelian" is the same with 
that of non-Abelian vortices \cite{Hanany:2003hp,Auzzi:2003fs,
Shifman:2004dr,Eto:2004rz,Eto:2005yh,
Balachandran:2005ev} 
carrying non-Abelian ${\mathbb C}P^{N-1}$ moduli;
see Refs.~\cite{Tong:2005un,Eto:2006pg,Shifman:2007ce,
Eto:2013hoa} for 
a review.
While  
a non-Abelian vortex 
can terminate on a non-Abelian monopole 
because of the matching of the moduli ${\mathbb C}P^{N-1}$  
\cite{Auzzi:2003em,Eto:2006dx},
a non-Abelian sine-Gordon soliton can terminate 
on a non-Abelian global vortex 
\cite{Balachandran:2002je,Nitta:2007dp,
Nakano:2007dq,Eto:2009wu,Eto:2013hoa}.
Non-Abelian sine-Gordon solitons 
exist stably in the 
color-flavor locking (CFL) phase of 
dense quark matter \cite{Alford:2001dt} 
or the confining phase of QCD 
as far as the axial anomaly term can be neglected 
at high density or high temperature 
\cite{Eto:2013bxa}.

In this paper, we construct the effective theory on 
the non-Abelian sine-Gordon soliton
by using the moduli approximation 
\cite{Manton:1981mp,Eto:2006uw},
that is a nonlinear sigma model 
with the target space ${\mathbb R} \times {\mathbb C}P^{N-1}$.
We then show that 
${\mathbb C}P^{N-1}$ lumps on it 
represent $SU(N)$ Skyrmions in the bulk point of view.
This setting offers a physical realization of 
the rational map Ansatz for $SU(N)$ Skyrmions 
of the Donaldson type.
One of the interesting features is that Skyrmions 
can exist stably without the Skyrme term.
This fact is consistent with the 
the Derrick's scaling argument 
\cite{Derrick:1964ww}  
that implies a three-dimensional soliton in scalar field theories 
shrinks in the absence of the Skyrme term, 
because the sine-Gordon soliton has 
divergent energy proportional to the world-volume directions.
This situation is similar to lumps inside a vortex 
representing a Yang-Mills instanton in the Higgs phase 
\cite{Eto:2004rz,Fujimori:2008ee}.

This paper is organized as follows.
In Sec.~\ref{sec:model}, 
we give  the $U(N)$ chiral Lagrangian 
and construct a non-Abelian sine-Gordon soliton.
In Sec.~\ref{sec:eff-th}, we construct 
the effective field theory on a non-Abelian 
sine-Gordon soliton which is 
the ${\mathbb C}P^{N-1}$ model.
In Sec.~\ref{sec:rational}, 
we show that 
${\mathbb C}P^{N-1}$ lumps on
the non-Abelian 
sine-Gordon soliton 
are nothing but $SU(N)$ Skyrmions in the bulk point of view.
Sec.~\ref{sec:summary} is devoted to 
summary and discussion.
In the Appendix, we summarize the Abelian sine-Gordon soliton.

\section{
$U(N)$ chiral Lagrangian 
and Non-Abelian sine-Gordon soliton
\label{sec:model}}
In this section,  we give 
the Lagrangian for a $U(N)$ principal chiral model (chiral Lagrangian) and 
its sine-Gordon solution.
A $U(N)$-valued field $U(x)$ takes a value in 
the $U(N)$ group having a nontrivial first homotopy group:
\beq
&& U(x) \in U(N) \simeq {U(1) \times SU(N) \over {\mathbb Z}_N} ,
\quad
 \pi_1 [U(N)] = {\mathbb Z}.
\eeq
The Lagrangian for a $U(N)$ chiral Lagrangian 
with the usual pion mass term is given by
\beq
{\cal L}/f_\pi^2 &=& \1{2} \tr \del_{\mu} U^\dagger \del^{\mu} U - {m^2 \over 2} \tr (2{\bf 1}_N - U - U^\dagger) \non
&=& \1{2} \tr (i U^\dagger \del_{\mu} U)^2 - {m^2 \over 2}\tr (2{\bf 1}_N - U - U^\dagger), \label{eq:U(N)SG}
\eeq
with $f_\pi$ being a constant of the mass dimension 1, 
and $\mu=0,1,\cdots,d-1$.
In the absence of the pion mass, $m=0$,
this Lagrangian is invariant under the chiral 
$SU(N)_{\rm L} \times SU(N)_{\rm R}$ symmetry 
\beq
 U(x) \to V_{\rm L}U(x)V_{\rm R}^\dagger , 
\quad V_{\rm L,R} \in SU(N)_{\rm L,R} \label{eq:U(N)sym0}
\eeq
that is spontaneously broken to the vectorlike symmetry 
\beq
 U(x) \to V U(x)V^\dagger , 
\quad V \in SU(N)_{\rm L+R=V} .\label{eq:U(N)sym}
\eeq
In the presence of the pion mass, $m\neq0$, 
the chiral symmetry is explicitly broken to 
the vectorlike symmetry in Eq.~(\ref{eq:U(N)sym}) 
in the unique vacuum $U={\bf 1}_N$.

The energy density for static configuration and 
its Bogomol'nyi completion are given as 
\beq
{\cal E}/f_\pi^2
&=& \1{2} \tr (i U^\dagger \del_x U)^2 
     - {m^2\over 2} \tr (2{\bf 1}_N - U - U^\dagger)\non
&=& \1{2} \tr \left[- {i\over 2} (U^\dagger \del_x U -  \del_x U^\dagger U ) \mp  m \sqrt {2 {\bf 1}_N - U -U^\dagger} \right]^2\non
&& 
\pm {m \over 2} \tr \left[ - {i\over 2} (U^\dagger \del_x U -  \del_x U^\dagger U ) \sqrt {2 {\bf 1}_N - U - U^\dagger)}\right] \non
&\geq& |t_{U(N)}| ,
\eeq
with the topological charge density defined by
\beq
t_{U(N)} \equiv
- {m \over 2} \tr \left[ i (U^\dagger \del_x U -  \del_x U^\dagger U ) \sqrt {2{\bf 1}_N - U- U^\dagger}\right].
\eeq
The BPS equation is obtained as
\beq
 - {i\over 2} (U^\dagger \del_x U -  \del_x U^\dagger U ) \mp  m \sqrt {2 {\bf 1}_N - U - U^\dagger}
={\bf 0}_N. \label{eq:BPS-U(N)}
\eeq
This equation is invariant under the $SU(N)_{\rm V}$ symmetry in Eq.~(\ref{eq:U(N)sym}).

A non-Abelian sine-Gordon soliton 
solution is of the following form:
\beq
 U(x) = V{\rm diag} (u(x),1,\cdots,1) V^\dagger , 
  \quad V\in SU(N)_{\rm V},  \label{eq:solution}
\eeq
with $u(x)$ ($u \in U(1), \; |u|^2=1$) satisfying 
the Abelian sine-Gordon equation
\beq
 -{i \over 2} (u^* \del_x u - (\del_x u^*) u )
 \mp m \sqrt {2-u-u^*} = 0 
\label{eq:BPS-U(1)}
\eeq
that allows for instance 
a single sine-Gordon soliton solution  \cite{Perring:1962vs}
\beq
 u(x) = \exp \left(4 i \, \arctan \exp [m  (x- X)] \right) 
\label{eq:U(1)-one-kink}
\eeq
with the boundary condition $u \to 1$ for $x \to \pm \infty$ (see Appendix).
Since there exists a redundancy in the action of $V$ 
in Eq.~(\ref{eq:solution}), 
$V$ in fact takes a value in the coset space
\beq
 V \in {SU(N)_{\rm V} \over SU(N-1)_{\rm V} \times U(1)_{\rm V}} \simeq {\mathbb C}P^{N-1}.
\eeq
The single-soliton solution has the moduli 
\beq
{\cal M} = {\mathbb R} \times {\mathbb C}P^{N-1},
\eeq
where the first and second factors are parametrized by 
$X$ and $V$, respectively.
In terms of the group elements, 
the general solution can be rewritten as
\beq
 U(x)
&=& \exp \left(i {\theta(x) \over N}\right) 
       \exp \left(i \theta(x)  VT_0 V^\dagger\right) \non
&=&  \exp \left(i {\theta(x) \over N}\right) \exp i{\theta(x)\over N}T \non
&=& \exp \left({i\theta(x) \phi\phi^\dagger}\right),
\label{eq:general} 
\eeq
with 
$T_0 \equiv {1 \over N} {\rm diag.}(N-1,-1,\cdots,-1)$, 
where 
$T \equiv V T_0 V^\dagger$ can be any $SU(N)$ generator normalized as $e^{i2\pi T} = \omega^{-1}{\bf 1}_N$ 
($\omega =\exp(2\pi i/N)$).
In the last line, we have introduced 
the orientational vector $\phi \in {\mathbb C}^N$
that represents homogeneous coordinates of 
${\mathbb C}P^{N-1}$ 
and satisfies
\beq
 && \phi^\dagger \phi =1,\label{eq:cond}\\
 && T = V T_0 V^\dagger = \phi \phi^\dagger  - {1\over N}{\bf 1}_N. \label{eq:phi}
\eeq
This form of $T$ is known as the 
projector in the rational map Ansatz for Skyrmions 
\cite{Houghton:1997kg,Ioannidou:1999mf},
already  implying the possibility of physical realization 
of the rational map.

\section{The Effective Theory on 
Non-Abelian Sine-Gordon Soliton}\label{sec:eff-th}

In this section, we 
construct the low-energy effective theory, 
which is the ${\mathbb C}P^{N-1}$ model,
by using the moduli approximation \cite{Manton:1981mp}.
Let us place a single sine-Gordon soliton perpendicular to 
the $x^3$-coordinate, that we denote $x$ for simplicity.
In the following, we will promote the moduli parameters $X$ and $\phi$ to be the fields on 
the $(2+1)$-dimensional soliton's world volume as
\beq
X \to X(x^\alpha),\quad \phi \to \phi(x^\alpha),\qquad
(\alpha = 0,1,2).
\eeq
We will derive the effective theory including 
derivatives with respect to $x^\alpha$ up to 
the leading (second) order,
by taking into account only the zero modes $X$ and $\phi$ and discarding massive modes. 
Therefore, what we will do in the rest of this section is integrating the kinetic
term of the chiral Lagrangian over $x$
\beq
{\cal L}_{\rm eff} = - \frac{f_\pi^2}{2} \int^\infty_{-\infty} dx\ \tr
\left[\left(U^\dagger \p_\alpha U\right)^2\right],
\label{eq:lag_eff}
\eeq
where $U$ is a non-Abelian sine-Gordon soliton solution in which the moduli parameters $X$ and $\phi$ are 
promoted to the fields on the world volume.
The effective Lagrangian correctly describe low energy physics with momenta sufficiently lower than 
the mass scale: $|p_\alpha| \ll m$.

\subsection{The $U(2)$ case}

As an exercise, we first consider the simplest case 
of $N=2$.  
We start with specifying an inhomogeneous coordinate $\varphi$ of the $\mathbb{C}P^1$ manifold
instead of the complex two-vector $\phi$ defined in  Eq.~(\ref{eq:phi}).
Note that $T$ defined in Eq.~(\ref{eq:phi}) is invariant under the $U(1)_{\rm V} \in SU(2)_{\rm V}$ transformation
\beq
T \to V V_0(\eta) T_0 V_0(\eta)^\dagger V^\dagger,\quad V_0(\eta) \equiv e^{i\eta T_0},
\eeq
with $\eta$ being an arbitrary real number. Therefore, an $SU(2)_{\rm V}$ matrix $V$ can 
always be transformed as $V \to V V_0(\eta)$. By using this $U(1)_{\rm V}$ transformation, one can
always cast the diagonal element of $V$ to be real valued. So, we will take the following 
concrete matrix
\beq
V = \frac{1}{\sqrt{1 + |\varphi|^2}}
\left(
\begin{array}{cc}
1 & -\varphi^* \\
\varphi & 1
\end{array}
\right),\quad \varphi \in \mathbb{C}.
\eeq
Then, we have 
\beq
T = \frac{1}{2(1+|\varphi|^2)} 
\left(
\begin{array}{cc}
1-|\varphi|^2 & 2\varphi^* \\
2\varphi & -(1-|\varphi|^2)
\end{array}
\right).
\eeq
The relation between $\phi$ and $\varphi$ can be found through 
the equation $T = \phi \phi^\dagger - {\bf 1}_2/2$ by 
\beq
\phi = V \left(
\begin{array}{c}
1\\
0
\end{array}
\right)
= \frac{1}{\sqrt{1 + |\varphi|^2}}
\left(
\begin{array}{c}
1\\
\varphi
\end{array}
\right).
\eeq
With these matrices, the concrete form of the matrix field  $U$ given in Eq. (\ref{eq:solution}) is
given by 
\beq
U(x;x^\alpha) = \frac{1}{1+|\varphi(x^\alpha)|^2} 
\left(
\begin{array}{cc}
u(x;X(x^\alpha)) + |\varphi(x^\alpha)|^2 & -(u(x;X(x^\alpha))-1)\varphi^*(x^\alpha)\\
-(u(x;X(x^\alpha))-1)\varphi(x^\alpha) & 1 + u(x;X(x^\alpha))|\varphi(x^\alpha)|^2
\end{array}
\right).
\eeq
Plugging this into Eq.(\ref{eq:lag_eff}), we have
\beq
{\cal L}_{\rm eff} = C_X \p_\alpha X \p^\alpha X 
+ C_\varphi \frac{\p_\alpha\varphi \p^\alpha \varphi^*}{\left( 1 + |\varphi|^2\right)^2},
\eeq
with
\beq
C_X &=& \frac{f_\pi^2}{2} \int^\infty_{-\infty} dx\ \left(\frac{\p\theta(x;X(x^\alpha))}{\p x}\right)^2 = \frac{f_\pi^2 T_{\rm sG}}{2},
\label{eq:eff_coeff1}\\
C_\varphi &=& 4 f_\pi^2 \int^\infty_{-\infty} dx\ \sin^2\frac{\theta(x;X(x^\alpha))}{2} = \frac{f_\pi^2 T_{\rm sG}}{m^2},
\label{eq:eff_coeff2}
\eeq
where $\theta$ is an ordinary sine-Gordon field which is related with $u$ by $u= e^{i\theta}$,
see the Appendix \ref{sec:sG}.
In the calculation above, 
we have used the BPS equation $\p_x \theta = \pm 2m \sin \theta/2$, and 
the tension of the sine-Gordon domain wall is given by
\beq
T_{\rm sG} = 8m.
\eeq

Some comments are in order: First, the coefficient $C_X = T_{\rm sG}/2$ of the translational zero
mode $X$ is consistent with the Nambu-Goto action of the order ${\cal O}(\p_\alpha^2)$.
Second, it is remarkable that 
the coefficient $C_\varphi$, 
called the K\"ahler class, has been exactly obtained.
The situation is similar to the BPS non-Abelian local vortex \cite{Hanany:2003hp,Shifman:2004dr}. 
Note that the K\"ahler class of 
the non-Abelian orientational zero modes cannot always be obtained. For example, 
the one for the non-BPS non-Abelian vortex in the dense QCD \cite{Balachandran:2005ev}
is only numerically determined.

\subsection{The $U(N)$ case} 

Now we generalize the results in the previous subsection for $N=2$ to the generic $N$.
Let us first specify the orientational zero modes as in the previous subsection.
Let $V_{ij}$ be an $(i,j)$ element of an $SU(N)_{\rm V}$ matrix. Since the $SU(N)$ generator $T_0$
is expressed as $(T_0)_{ij} = \delta_{i1}\delta_{j1} - \delta_{ij}/N$, Eq.~(\ref{eq:phi}) can be
written as
\beq
T_{il} = V_{ij}\left( \delta_{j1}\delta_{k1} - \delta_{jk}\frac{1}{N} \right) V_{lk}^* 
= V_{i1}V_{l1}^* - \delta_{il}\frac{1}{N} = \phi_i\phi_l^* - \delta_{il}\frac{1}{N}.
\eeq
We thus can identify $\phi$ as the first column vector of $V$, namely $\phi_i \equiv V_{i1}$.
Of course, the condition Eq.~(\ref{eq:cond}) is automatically satisfied:
$\phi^\dagger \phi = \phi_i^*\phi_i = V_{i1}^*V_{i1} = \delta_{11} = 1$.
Similarly, we can explicitly write down the matrix $U$ in Eq.~(\ref{eq:solution}) as
\beq
U_{il} = (VU_0V^\dagger)_{il} = V_{ij}\left(\delta_{jk} + (u-1)\delta_{1j}\delta_{1k}\right)V_{lk}^*
= \delta_{il} + (u-1) \phi_i\phi_l^*,
\eeq
where we have introduced
$U_0 = {\rm diag}(u,1,\cdots,1) \in U(N)$.
In the matrix notation, this can be simply expressed as
\beq
U = {\bf 1}_N + (u-1)\phi\phi^\dagger.
\eeq
Note that this can also  be derived from Eq.~(\ref{eq:general}) as
\beq
\exp\left(i\theta\phi\phi^\dagger\right)
&=& {\bf 1}_N + i \theta \phi \phi^\dagger 
+ \frac{1}{2!} \left(i \theta \phi \phi^\dagger\right)^2 
+ \frac{1}{3!} \left(i \theta \phi \phi^\dagger\right)^3 + \cdots \non
&=& {\bf 1}_N + \left(i\theta + \frac{1}{2!}(i\theta)^2 + \frac{1}{3!}(i\theta)^3 + \cdots\right) \phi\phi^\dagger \non
&=& {\bf 1}_N + \left(e^{i\theta} - 1\right) \phi\phi^\dagger.
\eeq

Thus, we have
\beq
\p_\alpha U = \phi\phi^\dagger \p_\alpha u 
+ (u-1)\left(\p_\alpha \phi\phi^\dagger + \phi\p_\alpha\phi^\dagger\right).
\eeq
By plugging this into the integrand of Eq.~(\ref{eq:lag_eff}), we find
\beq
\tr\left[\p_\alpha U \p^\alpha U^\dagger\right] =  \p_\alpha u \p^\alpha u + 
2 |1-u|^2 \left[\p_\alpha \phi^\dagger \p^\alpha \phi + (\phi^\dagger\p_\alpha \phi)(\phi^\dagger\p^\alpha \phi)
\right].
\eeq
In order to compute this, let us recall the following equations:
\beq
\p_\alpha u &=& \p_\alpha e^{i\theta(x;X(x^\alpha))} = i \p_\alpha X \frac{\p \theta}{\p X} u = - i \p_\alpha X \frac{\p\theta}{\p x} u,\\
2|1-u|^2 &=& 2(2 - u - u^*) = 8\sin^2 \frac{\theta}{2}.
\eeq
By plugging these into Eq.~(\ref{eq:lag_eff}) and performing the integral over $x$, we
again find the same integrals in Eqs.~(\ref{eq:eff_coeff1}) and (\ref{eq:eff_coeff2}).
Thus, we eventually reach the following Lagrangian for the generic $N$:
\beq
{\cal L}_{\rm eff} = \frac{f_\pi^2 T_{\rm sG}}{2} \p_\alpha X \p^\alpha X + \frac{f_\pi^2 T_{\rm sG}}{m^2}
\left[
\p_\alpha \phi^\dagger \p^\alpha \phi + (\phi^\dagger\p_\alpha \phi)(\phi^\dagger\p^\alpha \phi)
\right].
\eeq
The first term corresponds to the translational zero modes while the second term 
is the well-known Lagrangian for the $\mathbb{C}P^{N-1}$ nonlinear sigma model.

If one wants to express the $\mathbb{C}P^{N-1}$ Lagrangian in terms of the inhomogeneous
coordinate $\varphi^a$ ($a=1,2,\cdots,N-1$),
as in the previous subsection,
we take the $SU(N)_{\rm V}$ matrix
\beq
V = \frac{1}{\sqrt{1+|\vec\varphi|^2}}
\left(
\begin{array}{ccccc}
1 & - \varphi_1^* & - \varphi_2^* & \cdots & - \varphi_{N-1}^* \\
\varphi_1 & 1 + i \frac{|\vec\varphi|^2-|\varphi_1|^2}{|\vec\varphi|} & - i \frac{\varphi_1\varphi_2^*}{|\vec\varphi|} & \cdots  & - i \frac{\varphi_1\varphi_{N-1}^*}{|\vec\varphi|}\\
\varphi_2 & - i \frac{\varphi_2\varphi_1^*}{|\vec\varphi|} & \ddots 
& \ddots  & \vdots \\
\vdots &\vdots & \ddots & \ddots  & - i \frac{\varphi_{N-2}\varphi_{N-1}^*}{|\vec\varphi|} 
\\
\varphi_{N-1} & -i\frac{\varphi_{N-1}\varphi_1^*}{|\vec\varphi|} & \cdots &  -i\frac{\varphi_{N-1}\varphi_{N-2}^*}{|\vec\varphi|}  &
1 + i \frac{|\vec\varphi|^2-|\varphi_{N-1}|^2}{|\vec\varphi|}
\end{array}
\right).
\eeq
This $V$ includes $N-1$ complex parameters $\vec\varphi^T = ( \varphi_1, \varphi_2, \cdots, \varphi_{N-1})$.
A compact form of the elements of $V$ is given by
\beq
V_{ij} = \frac{1}{\sqrt{1+|\vec\varphi|^2}}\left(
\delta_{ij} + i \frac{|\vec\varphi|^2\delta_{ij} - \varphi_{i-1}\varphi_{j-1}^*}{|\vec\varphi|}\right),\qquad
\varphi_0 \equiv -i |\vec\varphi|.
\eeq
By making use of the identity $\sum_{i=1}^{N}|\varphi_{i-1}|^2 = 2 |\vec\varphi|^2$,
it is straightforward to check that the condition $(VV^\dagger)_{ik} = V_{ij}V_{kj}^* = \delta_{ik}$ is
satisfied. 
The effective Lagrangian can be rewritten as
\beq
 && {\cal L}_{\rm eff} = 
 \frac{f_\pi^2 T_{\rm sG}}{2} \p_\alpha X \p^\alpha X +
\frac{f_\pi^2 T_{\rm sG}}{m^2} 
g_{ab^*} \del_{\alpha} \ph^a \del^\alpha \ph^{*b}  ,\non
&& g_{ab^*} = \frac{\delta_{ab^*} (1+|\vec \ph|^2) - \ph^b \ph^{*a}} {(1+|\vec \ph|^2)^2}= \del_a \del_{b^*} \log (1 + |\vec \ph|^2).
\eeq

\section{${\mathbb C}P^{N-1}$ lumps on Sine-Gordon soliton as 
$SU(N)$ Skyrmions}  \label{sec:rational}

In this section, 
we construct 
${\mathbb C}P^{N-1}$ lumps 
in the effective theory on 
the sine-Gordon soliton in $d=3+1$ dimensions, 
and then show 
that they represent $SU(N)$ Skyrmions. 
A similar case was found before 
in the $SU(2)$ principal chiral model 
and the Skyrme model  
with a potential term admitting 
two discrete vacua \cite{Nitta:2012wi,Nitta:2012rq,Gudnason:2014nba}.

By placing a single sine-Gordon soliton perpendicular to 
the $x^3$-coordinate.
the effective theory on it is defined in 
the $x^0$, $x^1$, $x^2$ coordinates as in the last section. 
Apart from the translational modulus $X$,
the energy of static configuration
 and this Bogomol'nyi completion 
are given by
\beq
E &=&  \frac{f_\pi^2 T_{\rm sG}}{m^2}
\int d^2x 
 g_{ab^*}   (\del_1 \ph^a \del_1 \ph^{*b} + \del_2 \ph^a \del_2 \ph^{*b})\non
&=& \frac{f_\pi^2 T_{\rm sG}}{m^2} \int d^2x  
g_{ab^*} (\del_1 \ph^a \pm i \del_2 \ph^a)
 (\del_1 \ph^{*b} \mp i \del_2 \ph^{*b})
\pm \frac{f_\pi^2 T_{\rm sG}}{m^2} \int d^2x  \epsilon^{mn}
i g_{ab^*} \del_m \ph^a  \del_n \ph^{*b} \non
&\geq& |Q| \label{eq:BPS-bound}
\eeq
with spatial indices $m,n=1,2$ on the world volume.
Here, $Q$ is the topological lump charge defined by
\beq
 Q \equiv  \frac{f_\pi^2 T_{\rm sG}}{m^2} \int d^2x  \epsilon^{mn}
i g_{ab^*} \del_m \ph^a  \del_n \ph^{*b} 
  =  \frac{f_\pi^2 T_{\rm sG}}{m^2} 2\pi k  = {16 \pi f_\pi^2 \over m} k 
\eeq
with $k \in \pi_2 ({\mathbb C}P^{N-1})$ being the topological lump number.
The topological lump charge is 
the pullback of the K\"ahler form on ${\mathbb C}P^{N-1}$.
In terms of homogeneous coordinates $\phi$, 
the lump charge $k$ can also be expressed by \cite{Din:1980jg}
\beq
k = \frac{i}{2\pi} \int dz d\bar z\ \tr\left(\left[\p_{\bar z} {\cal P},\p_z {\cal P}\right]{\cal P}\right),
\quad {\cal P} \equiv \phi\phi^\dagger. \label{eq:lump-charge}
\eeq 
Note that ${\cal P}$ is a projection operator ${\cal P}^2 = {\cal P}$.

The inequality of the Bogomol'nyi energy 
bound in Eq.~(\ref{eq:BPS-bound}) is saturated if and only if 
the BPS or anti-BPS lump equation, 
\beq
 \del_{\bar z}\ph^a =0, \quad {\rm or} \quad \del_z\ph^a =0,\quad (a=1,2,\cdots,N-1),
\eeq
is satisfied, 
where we have defined a complex coordinate by $z\equiv x^1+ix^2$.
Generic BPS solutions in terms of $\phi$ are given by a set of holomorphic function $\{P_i(z)\}$,
\beq
\phi^T = (\phi_1,\cdots,\phi_N) 
  = \1{\sqrt{ \sum_{i=1}^N|P_i(z)|^2}} (P_1(z),P_2(z),\cdots,P_N(z)).
\eeq
The lump charge $k$ in Eq.~(\ref{eq:lump-charge})
corresponds to the degree of the highest-order polynomial $P_i(z)$ \cite{Din:1980jg}.
For instance, 
a single BPS lump solution in the ${\mathbb C}P^{N-1}$ model is 
given by 
\beq
k=1:\quad P_1 = z - z_0,\ P_2 = a,\ P_{i\ge3} = 0,
\eeq
up to symmetry, where $a$ is a complex modulus 
representing the size ($|a|$) and the phase (arg $a$), and
$z_0$ is the position moduli which we will set to be zero in the following. 

Let us take the non-Abelian sine-Gordon solution $U$ whose moduli parameter  $\phi$ is 
replaced by the lump solution
\beq
U(z,\bar z,x^3) 
&=& \exp \left({i\theta(x^3) \phi(z,\bar z)\phi^\dagger(z,\bar z)}\right) \non
&=& {\bf 1}_N + \left(u(x^3)-1\right) \phi(z,\bar z) \phi^\dagger(z,\bar z). \label{eq:rational}
\eeq
As long as  the condition $\p_1,\p_2 \ll \p_3 \sim m$ holds, 
this is an approximate solution of the full equations of motion in $3+1$ dimensions. 
So we should keep the size moduli of the lump $|a|^{-1}$ to be smaller than $m$.
For a configuration with $|a|^{-1} \gtrsim m$, one should take into account 
higher derivative corrections to the effective action or solve the full equations of motion in $3+1$
dimensions without using the effective theory at all, which we do not work out in this paper.

\begin{figure}[t]
\begin{center}
\includegraphics[width=0.7\linewidth,keepaspectratio]{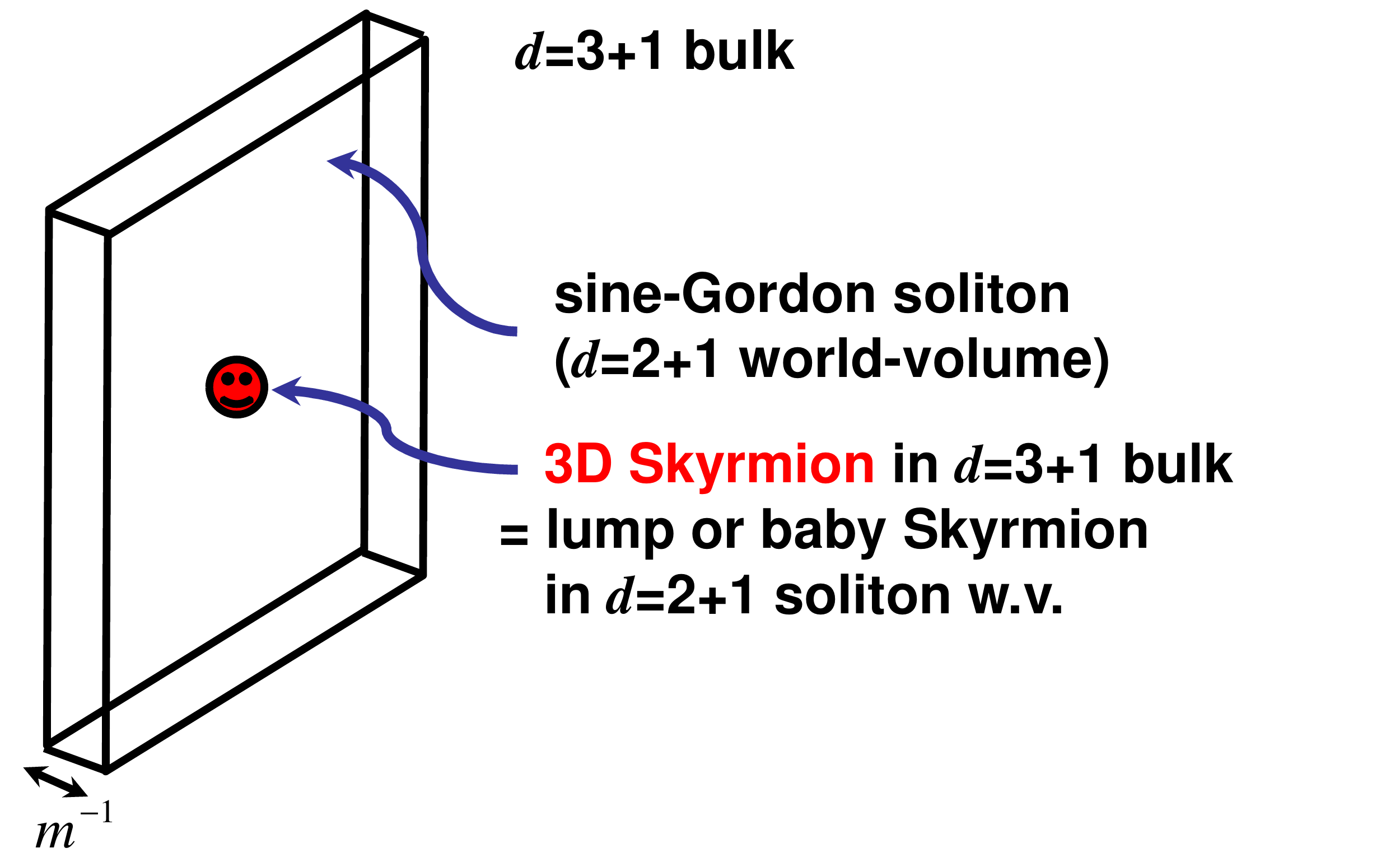}
\caption{
An $SU(N)$ Skyrmion as a ${\mathbb C}P^{N-1}$ lump inside 
a $U(N)$ non-Abelian sine-Gordon soliton.
\label{fig:lump-on-soliton}}
\end{center}
\end{figure}

\begin{figure}[t]
\begin{center}
\begin{tabular}{ccc}
\includegraphics[width=7cm]{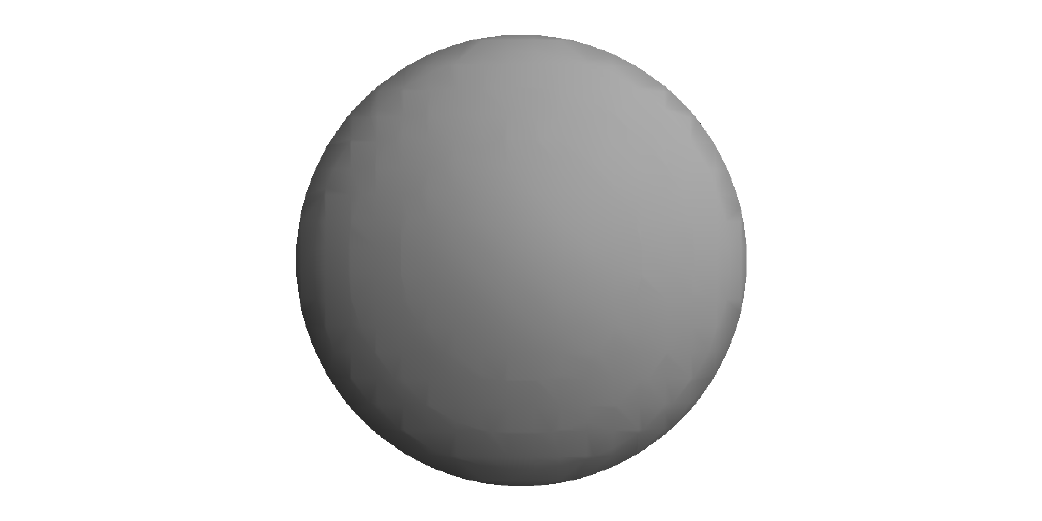}
&&
\includegraphics[width=7cm]{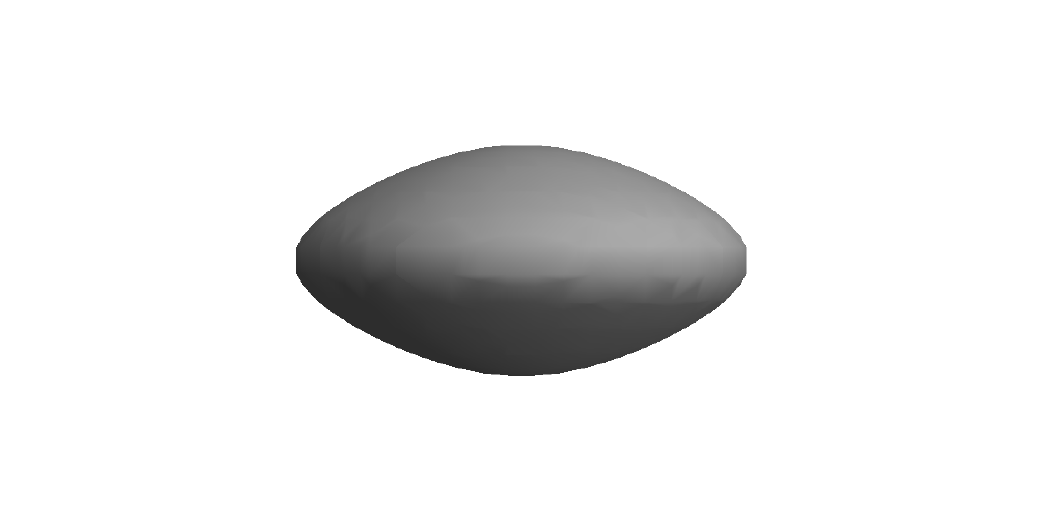}
\end{tabular}
\caption{An isosurface of the baryon density $(1/4\pi^2)\tr\left[R_1R_2R_3\right]$. We take $m=1$ and $|a|=2$.
The left panel shows the top view and the right panel shows the front view.}
\label{fig:Go_stone}
\end{center}
\end{figure}

By using the Maurer-Cartan one-form 
$R_i \equiv U^\dagger \del_i U$,
the baryon (Skyrmion) number 
$B$ taking a value in $\pi_3 [SU(N)]\simeq {\mathbb Z}$ 
in the bulk 
can be calculated as
\beq
B &=& \1{24 \pi^2} \int d^3x\  \epsilon_{ijk} \tr (R_iR_jR_k) \non
&=& -\1{8 \pi^2} \int  d^3x\  \tr \left[\left(\p_1U^\dagger\p_2U - \p_2U^\dagger\p_1U\right)  U^\dagger \p_3 U \right] \non
&=& -\frac{1}{8\pi^2}\int dx^1dx^2\ \tr\left(
\left[\p_1 {\cal P},\p_2{\cal P}\right] {\cal P}
\right) \int dx^3\ |u-1|^2 u^*\p_3u\non
&=&  \frac{1}{8\pi^2}\int dzd\bar z\ \tr\left(
\left[\p_{\bar z} {\cal P},\p_z{\cal P}\right] {\cal P}
\right) \int dx^3\ 4 i  \sin^2 \frac{\theta}{2} \p_3\theta\non 
&=&  \frac{i}{2\pi}\int dzd\bar z\ \tr\left(
\left[\p_{\bar z} {\cal P},\p_z{\cal P}\right] {\cal P}
\right) \times \frac{1}{2\pi} \int dx^3\  (1-\cos\theta) \p_3\theta\non
&=& k \times \frac{\Delta}{2\pi}
\eeq
where we have defined 
the sine-Gordon soliton charge by
\beq
\Delta \equiv \theta(x^3=+\infty) - \theta(x^3=-\infty) = 2\pi (n_+ - n_-),\qquad (n_\pm \in \mathbb{Z}).
\eeq
The single non-Abelian sine-Gordon soliton 
has $\Delta = 2\pi$.
Therefore, we have found that ${\mathbb C}P^{N-1}$ lumps
on the non-Abelian sine-Gordon soliton represent
$SU(N)$ Skyrmions 
as illustrated in Fig.~\ref{fig:lump-on-soliton}.
Note that the Skyrmion confined in the non-Abelian sine-Gordon soliton is not spherical but
looks like a pancake, a Go stone, or M$\&$M's candy; see Fig.~\ref{fig:Go_stone}.
Although this calculation is valid only for $a$ satisfying $|a| \gg m^{-1}$, 
the result is independent of the size moduli $a$. Therefore, we expect that
this is true for any $a$.

\begin{figure}[t]
\begin{center}
\begin{tabular}{ccccc}
\includegraphics[width=5cm]{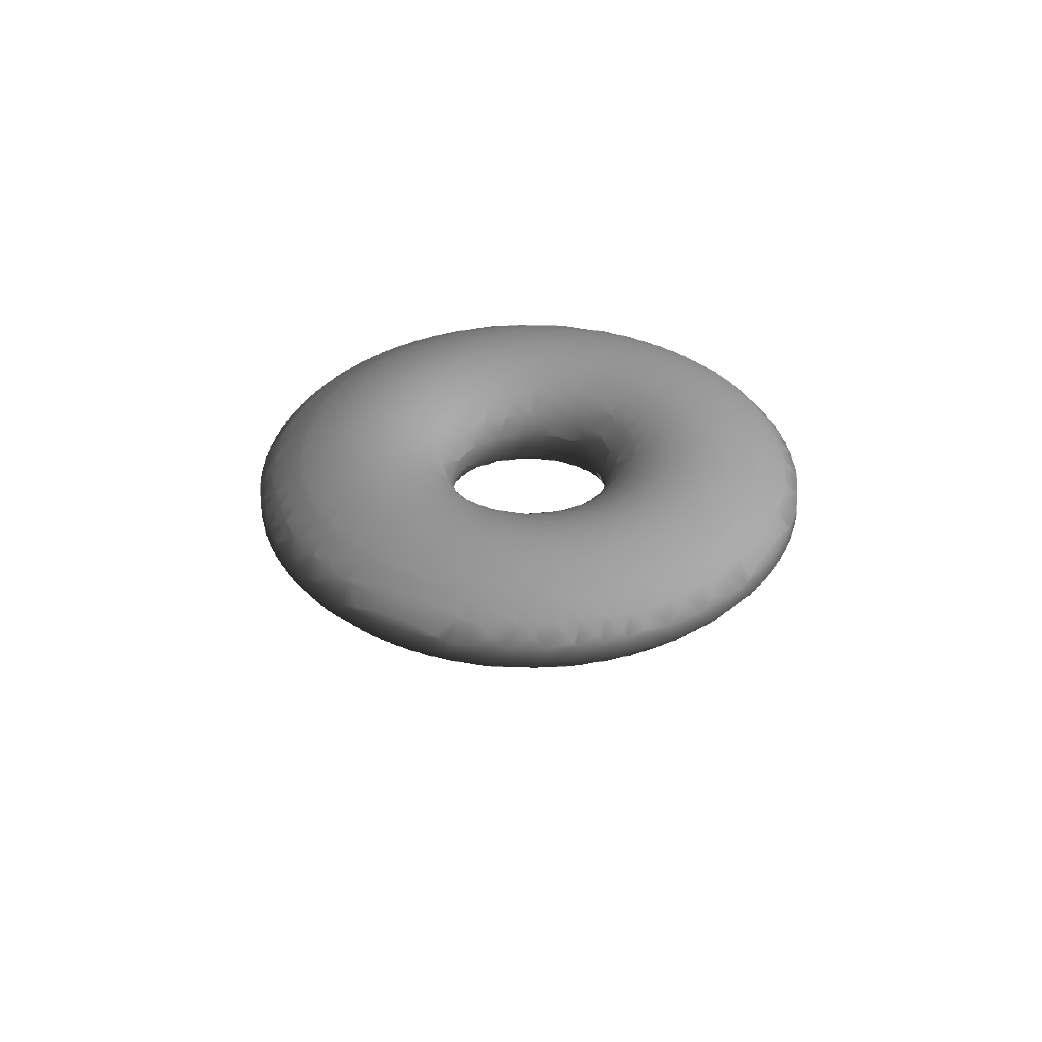} &&
\includegraphics[width=5cm]{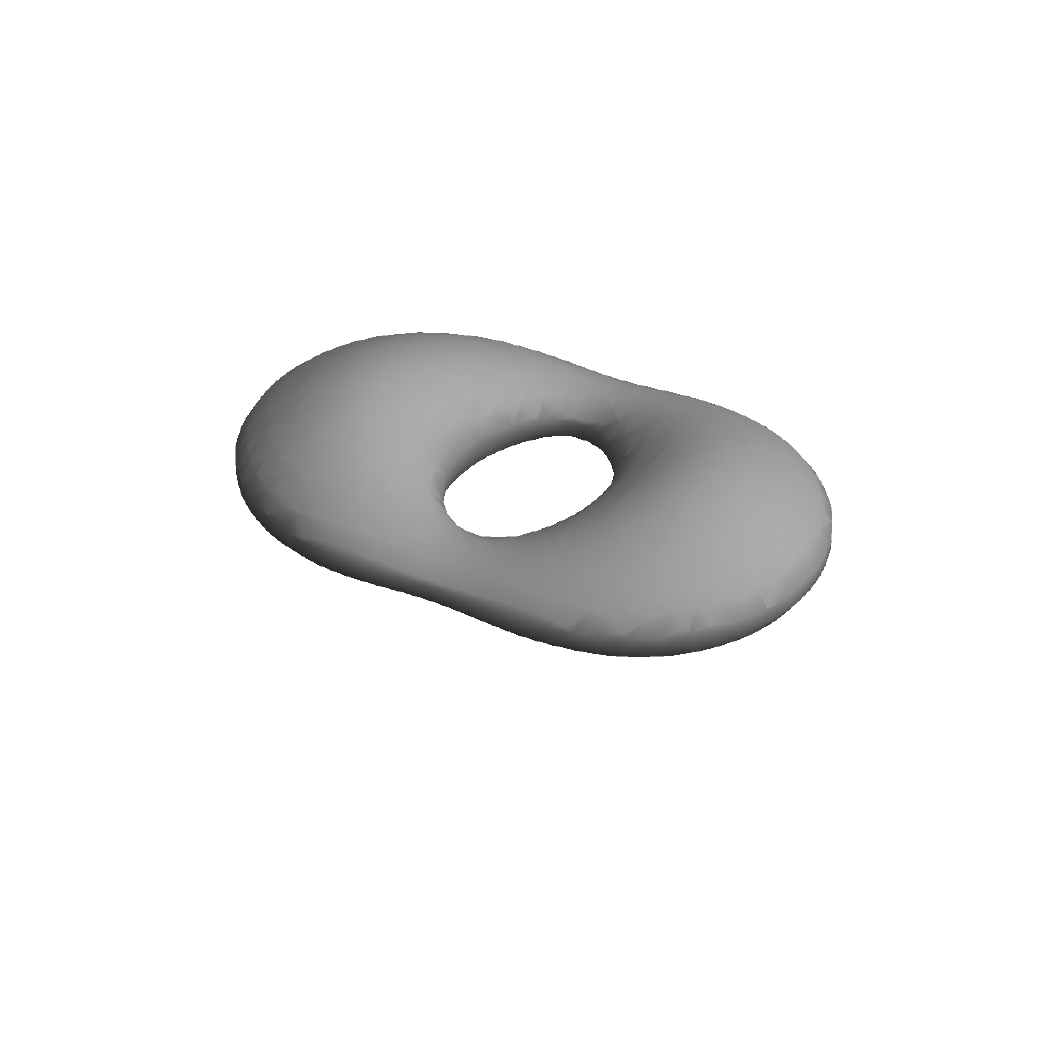} &&
\includegraphics[width=5cm]{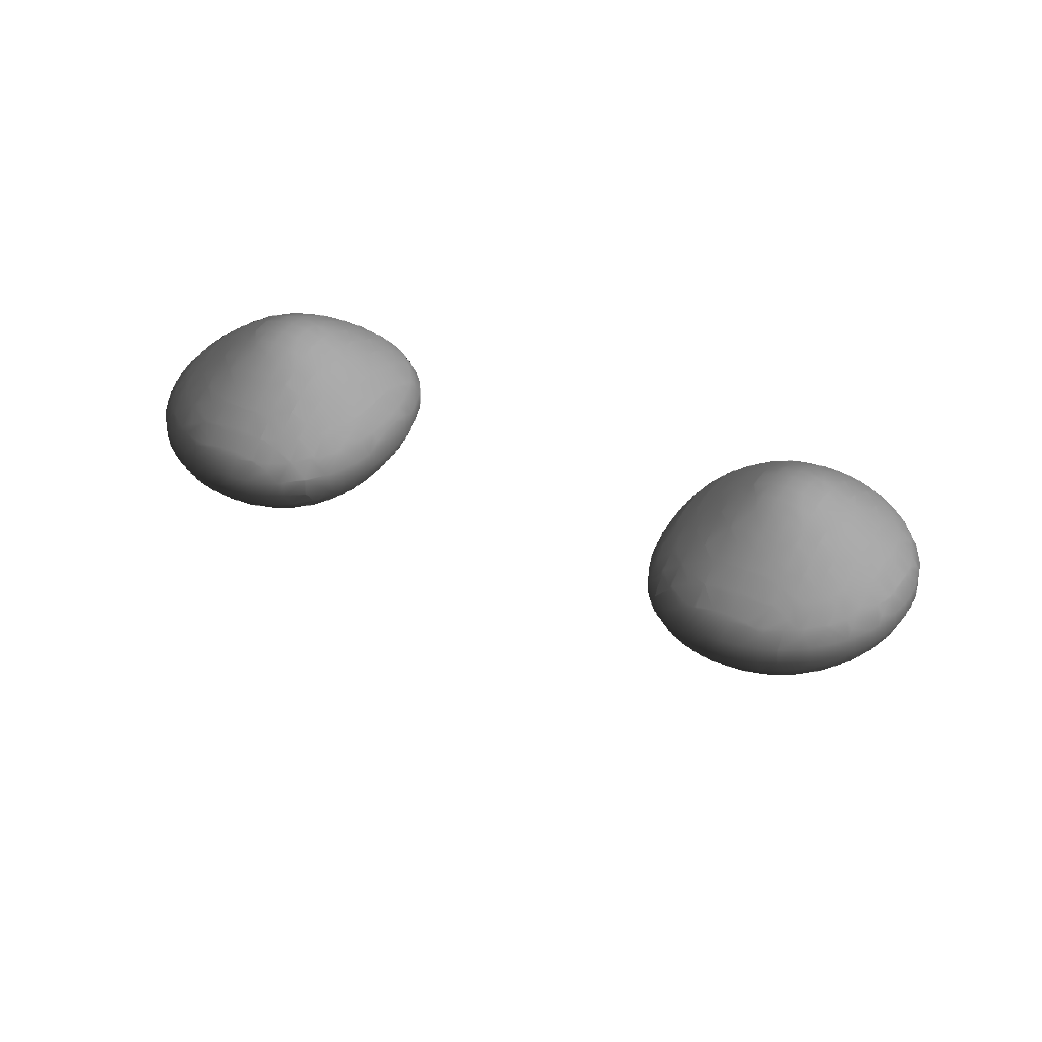}
\end{tabular}
\caption{$B=2$ Skyrmions: We fix $|a|=10$ and change $d$.
The left, middle, and right  panels have $d=0,2,4$, respectively.}
\label{fig:k2}
\end{center}
\end{figure}

\begin{figure}[t]
\begin{center}
\begin{tabular}{ccccc}
\includegraphics[width=5cm]{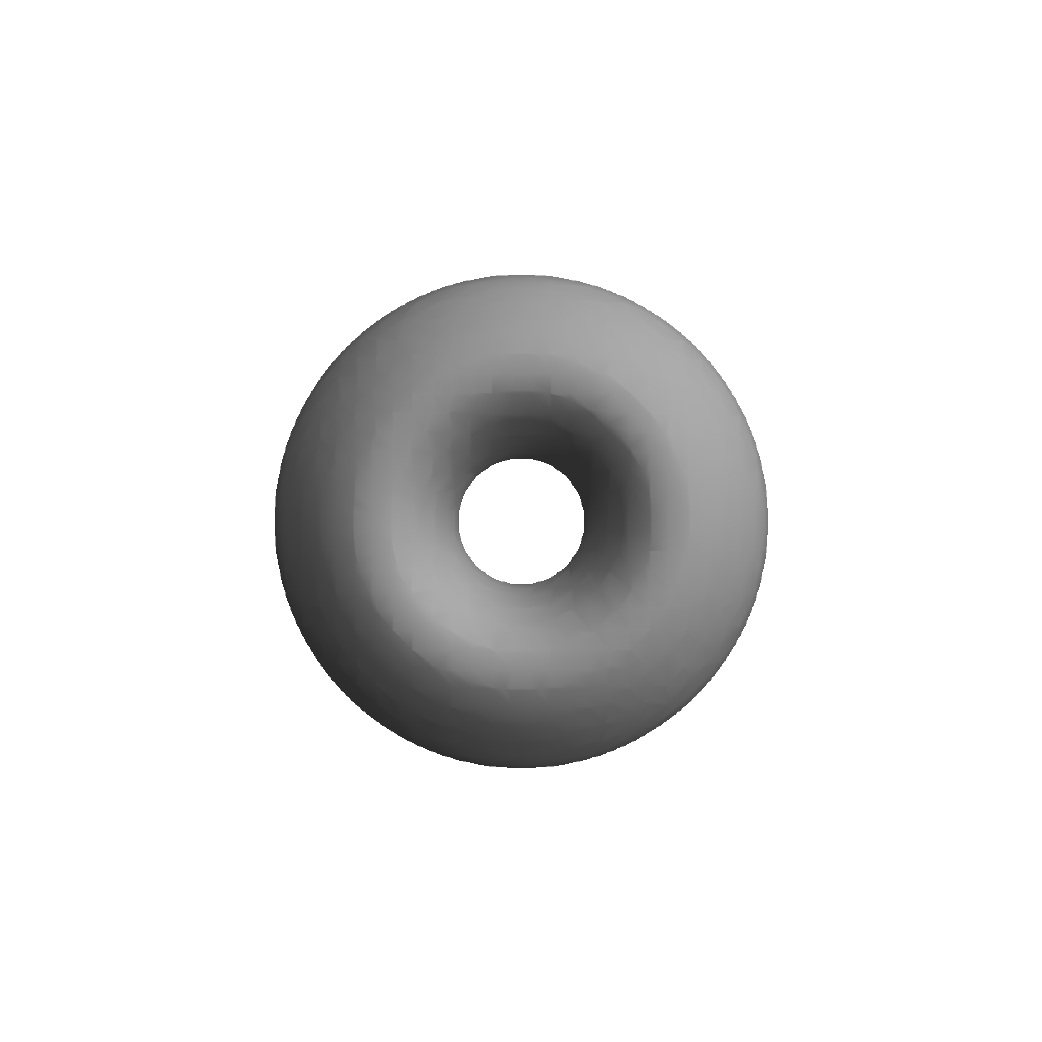} &&
\includegraphics[width=5cm]{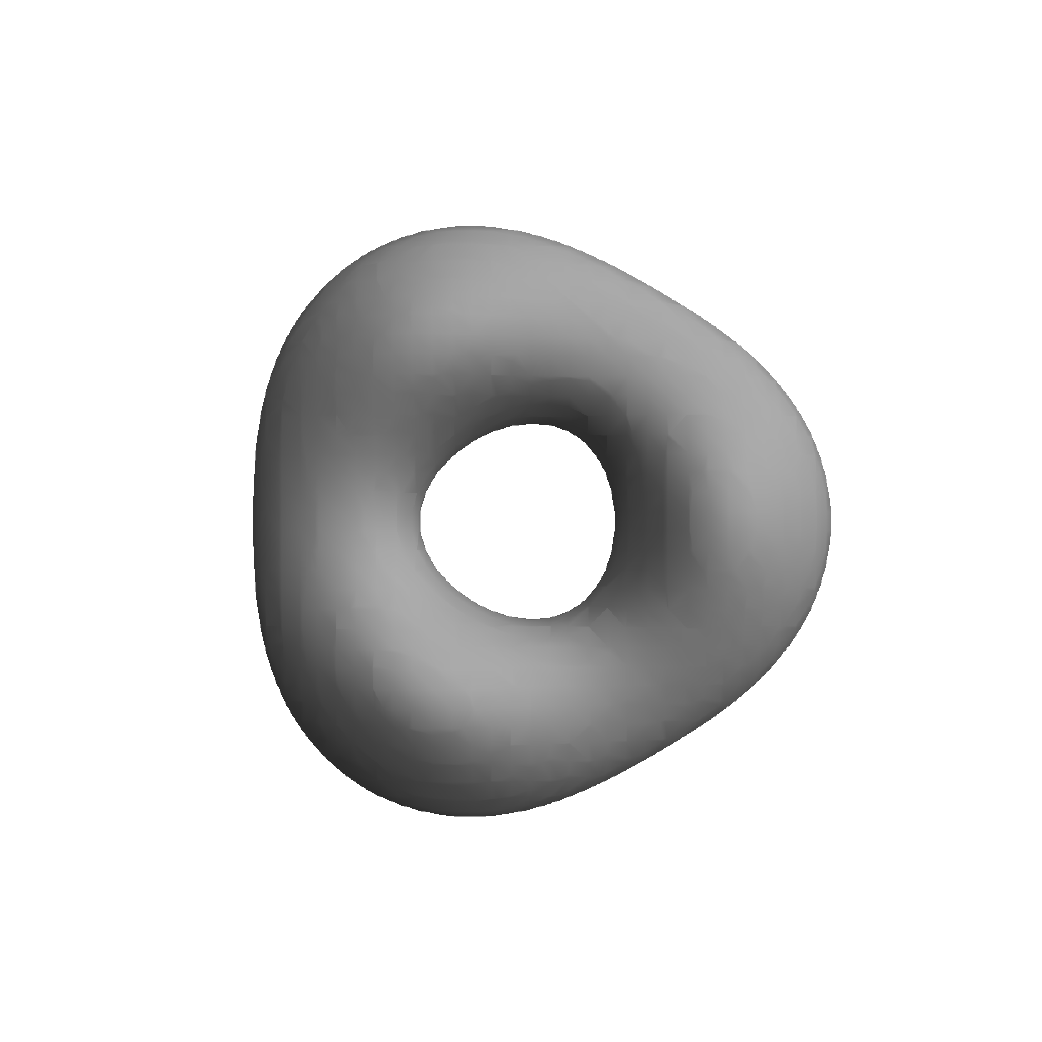} &&
\includegraphics[width=5cm]{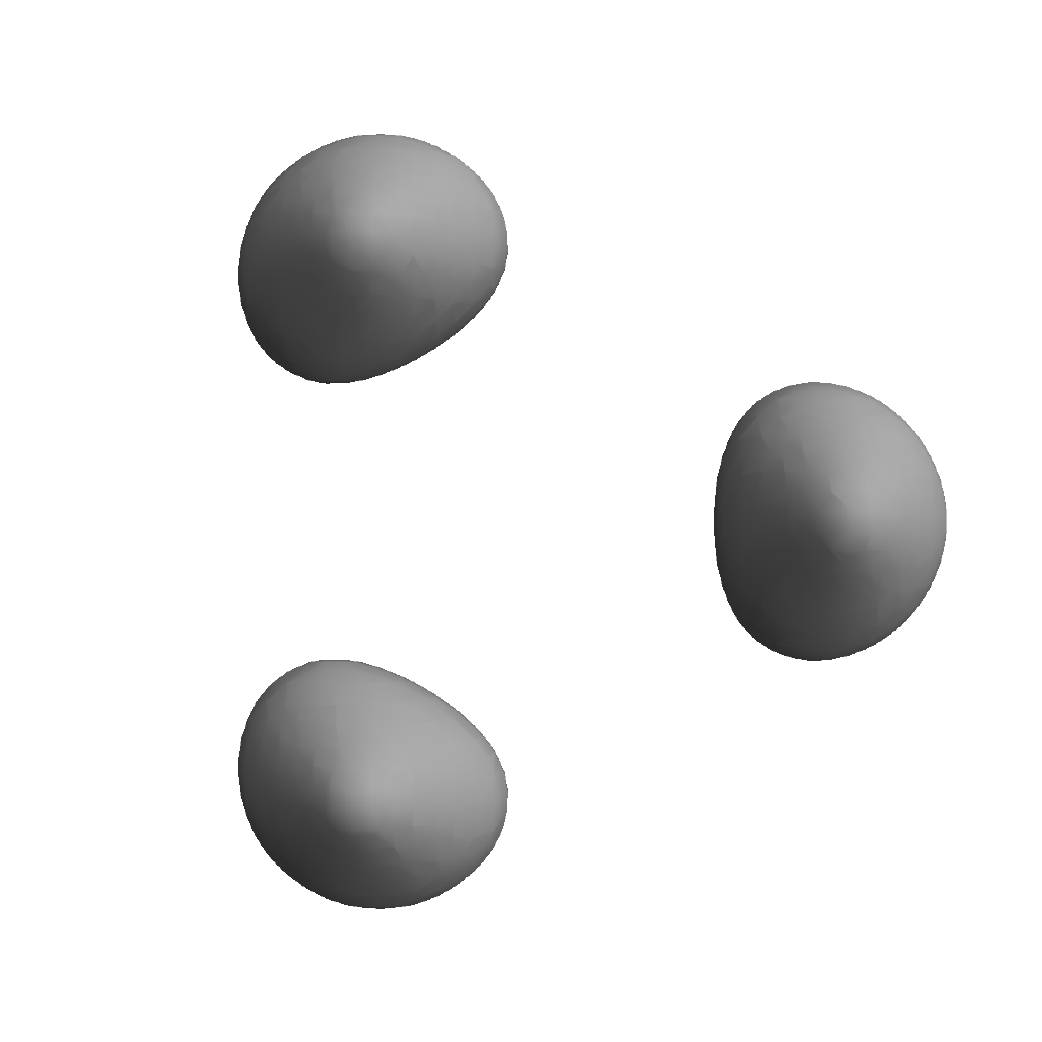}
\end{tabular}
\caption{$B=3$ Skyrmions: The left, middle, and right panels have
$(a,d) = (10,0), (20,2), (40,4)$, respectively.}
\label{fig:k3}
\end{center}
\end{figure}

Configurations with higher baryon charges are also easy to construct
in terms of the effective theory. For example, $B=2$ configurations are described by
\beq
k=2:\quad 
P_1=(z-d)(z+d),\ P_2=a,\ P_{k\ge3} = 0.
\eeq
Two Skyrmions sit at $z= \pm d$. Some configurations are shown in Fig.~\ref{fig:k2}. 
When the two Skyrmions have an overlap region, the baryon
density isosurface becomes a donut shape as usual $B=2$ Skyrmions in the Skryme model.
A difference appears for $k\ge 3$. It is known that when usual $B=3$ Skyrmions coincide on top of each other, the baryon density
exhibits a tetrahedral structure. On the other hand, since the Skyrmions are confined inside a soliton in our
model, such a three-dimensional structure does not appear. 
For instance, a ${\mathbb Z}_3$ symmetric 
$B=3$ configuration can be given by
\beq
k=3:\quad
P_1 = (z-d)(z-d\omega)(z-d\omega^2),\ P_2 = a,\ P_{i\ge3} = 0,
\eeq
with $z=1,\omega,\omega^2$ being roots of $z^3 =1$. 
As can be seen in Fig.~\ref{fig:k3}, instead of having a tetrahedron, a torus structure again appear when multiple
Skyrmions coincide ($d=0$) inside the sine-Gordon soliton.

We note that configurations in 
Eq.~(\ref{eq:rational}) show 
a physical realization of the rational map Ansatz
of the Donaldson type.
For conventional Skyrmions, 
rational maps give merely initial configurations for 
numerical relaxations, 
although 
a spherical version of the rational map eventually 
gives a good approximation to the final configurations 
\cite{Houghton:1997kg,Manton:2004tk,Ioannidou:1999mf,Lau:2014baa}.
On the other hand, 
we would like to emphasize that, 
in our case, 
the rational map of the Donaldson type solves
the equation of motion in the moduli approximation, 
that is, as far as 
the condition $\p_1,\p_2 \ll \p_3 \sim m$ holds.

We have seen that the Skyrmion can exist stably 
even in the absence of the Skyrme term. 
One might have concerns about
Derrick's theorem \cite{Derrick:1964ww} 
implying that Skyrmions should shrink 
without four (higher) derivative terms 
in three spatial dimensions in the bulk. 
However, Derrick's theorem  in 
the whole three dimensions cannot be applied to
our case because of the divergent energy 
of the sine-Gordon soliton linearly 
extended to two spacial directions. 
In the presence of such an extended object, 
Derricks's theorem should be applied to 
each direction separately; 
we first use Derrick's theorem in one dimension 
for the sine-Gordon soliton 
ensuring its stability.
Then we use it  for two dimensions 
in the soliton's world volume, 
ensuring the marginal stability of the lumps.
In addition, topology which supports our solution
is a combination of 
$\pi_1[U(N)] \simeq {\mathbb Z}$ for 
the sine-Gordon soliton
and $\pi_2({\mathbb C}P^{N-1}) \simeq {\mathbb Z}$ for the lumps.
This situation is parallel to lumps inside a vortex 
corresponding to Yang-Mills instantons in the Higgs phase
\cite{Eto:2004rz,Fujimori:2008ee}. 

The sine-Gordon soliton is BPS saturating the Bogomol'nyi bound 
and lumps are also 
BPS saturating the Bogomol'nyi bound  
in the world-volume theory.
However the Skyrmion as the composite soliton itself is not BPS.

\section{Summary and Discussion}\label{sec:summary}

We have constructed the effective theory on 
a non-Abelian sine-Gordon soliton 
in the $U(N)$ chiral Lagrangian 
to obtain 
the nonlinear sigma model 
with the target space ${\mathbb R} \times {\mathbb C}P^{N-1}$.
We have shown that 
${\mathbb C}P^{N-1}$ lumps on the  
non-Abelian sine-Gordon soliton 
are nothing but 
$SU(N)$ Skyrmions in the bulk point of view.
This setting offers a physical realization of 
the rational map Ansatz for Skyrmions 
of the Donaldson type 
that solves the equations of motion 
in the moduli approximation.
Skyrmions can exist stably 
inside the soliton without the Skyrme term.

Several discussions are addressed here. 
In this paper, we have used 
the moduli approximation to construct 
the Skyrmions 
trapped inside the sine-Gordon soliton.
Full numerical analyses of the stability 
beyond the moduli approximation 
are available in related models.
The stable solution in the $SU(2)$ principal chiral model 
with two vacua 
without any higher derivative terms 
was obtained numerically \cite{Gudnason:2014nba}. 
Furthermore, dynamical stability was 
tested in Ref.~\cite{Jennings:2013aea} 
for configurations similar to ours
in lower dimensions  
(baby Skyrmions trapped inside 
the domain wall in 2+1 dimensions).

Since we have not considered the Skyrme term in this paper, 
the ${\mathbb C}P^{N-1}$ Lagrangian 
on the soliton 
admits ${\mathbb C}P^{N-1}$  lumps with arbitrary sizes 
and there is no force between Skyrmions.
If we add the Skyrme term in the original 
Lagrangian, it will induce 
a baby-Skyrme term (as well as enhancement 
of the kinetic term \cite{Gudnason:2014gla,Gudnason:2014hsa}) in the ${\mathbb C}P^{N-1}$ model 
on the soliton, 
which was the case for 
the $SU(2)$ chiral Lagrangian with 
two discrete vacua  
\cite{Nitta:2012wi,Nitta:2012rq,Gudnason:2014nba,
Gudnason:2014hsa}.
In this case, the lumps are unstable to expand.
In order to stabilize them, one has to further 
introduce a mass term that explicitly breaks  
the $SU(N)_{\rm V}$ symmetry in the bulk, 
resulting in 
${\mathbb C}P^{N-1}$ baby Skyrmions 
\cite{Piette:1994ug}
on the soliton corresponding to
$SU(N)$ Skyrmions in the bulk.

In this paper, we have constructed the effective theory on a single soliton. 
It is well known that 
the sine-Gordon equation admits more general solutions 
such as a breather solution of two solitons and 
a static multiple soliton lattice.
Constructing the effective theory on these cases  
will be interesting while paying attention 
to the relation between orientational zero modes 
localized on different solitons, 
in particular for a soliton lattice.
The orientational modes 
on a non-Abelian $U(N)$ vortex lattice 
have been discussed recently, 
and found to give 
an inhomogeneous ${\mathbb C}P^{N-1}$ model 
\cite{Kobayashi:2013axa}.
Therefore, a similar mechanism may work here.

In the Skyrme model, Skyrmions are identified with baryons.
In the quark model, baryons are composite states of constituent quarks.
Hence, one would naively expect that Skyrmions are also composite sates of 
constituent solitons in the Skyrme model. Unfortunately,  no fractional solitons
that could be identified with quarks
have been found in the original Skyrme model of hadrons.
On the other hand,
there can exist fractional solitons 
 in our model 
with some modifications.
It is known that 
one $\mathbb{C}P^{N-1}$ lump can be
decomposed into 
$N$ fractional lumps (merons) with 
$1/N$ lump charges 
in certain situations  such as 
a twisted boundary condition
\cite{Eto:2004rz,Eto:2006mz,Bruckmann:2007zh},
an introduction of a suitable potential \cite{Schroers:1995he,Jaykka:2010bq} 
or 
a deformation of the target space metric 
\cite{Collie:2009iz,Eto:2009bz}.
Since the lump is identified with the Skyrmion (baryon) in our model, the merons with $1/N$ baryon charge
might be identified with quarks. 
The ``quarks"
are confined to baryons 
and cannot be observed in our model as it is. 
In order to obtain deconfined quarks, 
we might need to break 
the chiral symmetry explicitly. 
We will report this interesting problem elsewhere.

The CFL phase of 
dense quark matter \cite{Alford:2001dt} 
admits a non-Abelian $U(3)$ sine-Gordon soliton. 
Therefore, we can construct $SU(3)$ Skyrmions stably 
inside a non-Abelian sine-Gordon soliton.
It was conjectured that 
Skyrmions in the CFL phase are quarks (called qualitons)
rather than baryons as in the usual Skyrme model 
\cite{Hong:1999dk,Eto:2013hoa}.
In the CFL phase, however, it was a 
problem that Skyrmions cannot exist stably 
in the absence of the Skyrme term.
In our case, they exist stably 
inside a non-Abelian sine-Gordon soliton. 
Physical implications of this remain a future problem.

While a non-Abelian sine-Gordon soliton is stable in 
the framework of a chiral Lagrangian, 
it can be unstable or metastable 
in a linear sigma model 
because it can be terminated by  
a non-Abelian global vortex \cite{Nitta:2014rxa}.
Consequently, a soliton is bounded by 
a closed loop of a non-Abelian vortex. 
The effective theory is therefore 
the ${\mathbb C}P^{N-1}$ model with the boundary, 
which will be interesting to explore.

We have considered the group $U(N)$ for sine-Gordon solitons. 
In the case of non-Abelian vortices, 
$U(N)$ were extended to  
arbitrary gauge groups $G$ in the form of 
${G \times U(1) \over {\mathbb Z}_r}$ 
with the center ${\mathbb Z}_r$ of $G$ of rank $r$
 \cite{Eto:2008yi} 
 such as $SO(N)$ and $USp(2N)$ groups \cite{Eto:2009bg}.
In the same way,
non-Abelian $U(N)$ sine-Gordon solitons can also be
extended to such cases.
The effective theory on such a $G$ sine-Gordon soliton 
can be constructed to obtain 
a nonlinear sigma model with the target space 
${\mathbb R} \times G/H$ with a suitable subgroup $H$.
In this case, $G/H$ lumps on the sine-Gordon soliton 
will represent $G$ Skyrmions. 

The composite Skyrmions constructed in this paper 
are not BPS
although their constituents, sine-Gordon solitons and lumps, 
are all BPS. 
On the other hand, the Skyrmions are BPS in the BPS Skyrme model 
consisting of a six-derivative term and a potential term
\cite{Adam:2010fg}.
A corresponding model to our $U(N)$ case 
may admit BPS Skyrmions as BPS lumps inside a BPS soliton.

We have not  discussed supersymmetry in this paper.
If we promote the target space $U(N)$ to 
$T^* U(N) \simeq GL(N,{\mathbb C})$, 
the model can be made supersymmetric 
\cite{Nitta:2014fca}. 
For that case, sine-Gordon solitons and lumps 
may preserve a half supersymmetry, 
while the total configuration breaks all supersymmetry 
because it is non-BPS.

As a lower-dimensional analogue, 
a lump (baby Skyrmion) can be constructed 
\cite{Nitta:2012xq,Jennings:2013aea} 
as a sine-Gordon soliton on a ${\mathbb C}P^1$ kink 
\cite{Abraham:1992vb}. 
This relation can be generalized to 
${\mathbb C}P^{N-1}$ lumps as sine-Gordon solitons on ${\mathbb C}P^{N-1}$ kinks \cite{Gauntlett:2000ib}. 
Combined with the result in this paper, 
$SU(N)$ Skyrmions can be constructed 
only from domain walls, 
as was so for $SU(2)$ Skyrmions \cite{Nitta:2012rq}.

As shown in this paper,
the target space of the effective theory 
on a single non-Abelian sine-Gordon soliton 
is ${\mathbb C}P^{N-1}$ 
having the nontrivial second homotopy group 
$\pi_2 ({\mathbb C}P^{N-1}) \simeq {\mathbb Z}$.
In the latter part of the paper, we have constructed
the lump solutions as topological textures 
characterized by this homotopy group. 
On the other hand, the same homotopy group 
admits a monopole as a topological defect 
if the soliton world volume is 3+1 dimensional, 
that is, the bulk is 4+1 dimensional.
This gives a D-brane soliton, that is, 
a Skyrmion string ending on 
a domain wall, as has been recently 
shown in Ref.~\cite{Gudnason:2014uha}  
for the $SU(2)$ Skyrme model with two vacua, 
as a higher-dimensional generalization 
of lump strings ending on a domain wall 
in the ${\mathbb C}P^1$ model \cite{Gauntlett:2000de},  
${\mathbb C}P^N$ 
or Grassmann sigma model \cite{Isozumi:2004vg}.
One advantage of our model 
is the existence of 
parallel solitons as many as possible without antisolitons, 
in contrast to previous works 
(the ${\mathbb C}P^N$ model  admits at most $N$ parallel walls).

\section*{Acknowledgments}

This work is supported in part by JSPS KAKENHI Grant 
No. 23740198 (M.~E.) and 
No.~25400268 (M.~N.)
and by the ``Topological Quantum Phenomena'' 
Grant-in-Aid for Scientific Research 
on Innovative Areas (No.~25103720 (M.~N.))  
from the Ministry of Education, Culture, Sports, Science and Technology 
(MEXT) of Japan. 

\begin{appendix}
\section{The sine-Gordon model}
\label{sec:sG}

Here, we summarize the conventional sine-Gordon soliton 
to fix notations.
The Lagrangian density of the conventional sine-Gordon model is 
\beq
 {\cal L} &=& \1{2} (\del_{\mu} \theta)^2 
- m^2 \left(1-\cos \theta\right) 
 \label{eq:SG}
\eeq
with $\mu=0,1,\cdots,d-1$ and $0 \leq \theta <2\pi$.
The sine-Gordon soliton is characterized by
the first homotopy group $\pi_1[U(1)] \simeq {\mathbb Z}$.
The static energy density of static configurations depending on one spatial direction $x$ and its  Bogomol'nyi completion are given by
\beq
 {\cal E} &=& \1{2} (\del_x \theta)^2 
+ m^2 \left(1-\cos \theta\right) \non
&=& \1{2} (\del_x \theta)^2 + 2 m^2 \sin^2{\theta \over 2} \non
 &=& \1{2} \left(\del_x\theta \mp 2 m \sin {\theta\over 2}\right)^2 
  \pm 2 m \partial_x \theta \sin {\theta\over 2} \non
 &\geq& 
 \left|2 m \partial_x \theta \sin {\theta\over 2}  \right|
 = \left|t_{\rm sG}\right|
\eeq
with the topological charge density defined by
\beq
t_{\rm sG} 
 \equiv 2m \partial_x \theta \sin {\theta\over 2}
 = - 4 m \partial_x \left( \cos {\theta \over 2}\right).
\eeq
The inequality is saturated by the BPS equation
\beq
 \del_x \theta \mp 2 m \sin {\theta\over 2} = 0.
\eeq
A single-soliton solution interpolating between 
$\theta =0$ at $x \to -\infty$ to 
$\theta =2\pi$ at $x \to +\infty$ and 
its topological charge are
\beq
&& \theta(x) = 4 \arctan \exp{m (x- X)} ,\\
&& T_{\rm sG} = \int dx t_{\rm sG} 
= -4 m \left[\cos {\theta\over 2}\right]^{x=+\infty}_{x=-\infty}
= -4 m (-1-1) = 8 m,
\label{eq:SG-charge}
\eeq
respectively. Here, $X$ is the sine-Gordon soliton position and 
the width of the soliton is $1/m$.

By using the field $u \equiv e^{i\theta}$ and
taking a value in the $U(1)$ group, 
the BPS equation,  the topological charge density 
and the single-soliton solution 
can be rewritten as
\beq
&& -{i \over 2} (u^* \del_x u - (\del_x u^*) u )
 \mp m \sqrt {2-u-u^*} = 0 ,
\label{eq:BPS-U(1)}\\
&& t_{\rm U(1)} 
= -{i m\over 2} (u^* \del_x u - (\del_x u^*) u )
       \sqrt {2-u-u^*}
= - 2 m \del_x  \left( \sqrt {2+ u+u^*} \right) ,\\
&&
 u(x) = \exp \left(4 i \, \arctan \exp [m  (x- X)] \right) ,
\label{eq:U(1)-one-kink}
\eeq
respectively, with the boundary condition $u \to 1$ for $x \to \pm \infty$.

\end{appendix}

\newcommand{\J}[4]{{\sl #1} {\bf #2} (#3) #4}
\newcommand{\andJ}[3]{{\bf #1} (#2) #3}
\newcommand{\AP}{Ann.\ Phys.\ (N.Y.)}
\newcommand{\MPL}{Mod.\ Phys.\ Lett.}
\newcommand{\NP}{Nucl.\ Phys.}
\newcommand{\PL}{Phys.\ Lett.}
\newcommand{\PR}{ Phys.\ Rev.}
\newcommand{\PRL}{Phys.\ Rev.\ Lett.}
\newcommand{\PTP}{Prog.\ Theor.\ Phys.}
\newcommand{\hep}[1]{{\tt hep-th/{#1}}}

\end{document}